\documentstyle [11pt,menu97]{article}
\begin{document}
    \setlength{\baselineskip}{2.6ex}
\title{Nucleon-nucleon partial wave analysis to 2.5 GeV}
\author{Richard Arndt, Chang-Heon Oh, Igor Strakovsky, and Ron Workman\\
{\em Department of Physics, Virginia Polytechnic Institute and State
University\\
Blacksburg, VA 24061-0435}\\
\vspace{0.2cm}
and \\
\vspace{0.2cm}
Frank Dohrmann\\
{\em I. Institut f\"ur Experimentalphysik, Universit\"at Hamburg,
Luruper Chaussee 148 \\
D-22761 Hamburg, Germany}}
\maketitle
\begin{abstract}
\setlength{\baselineskip}{2.6ex}
A partial-wave analysis of $NN$ elastic scattering data has been
completed.  This analysis covers an expanded energy range, from
threshold to a laboratory kinetic energy of 2.5~GeV, in order to
include recent elastic $pp$ scattering data produced by the EDDA
collaboration at COSY.  The results of both single-energy and
energy-dependent analyses are described.
\end{abstract}

\section*{INTRODUCTION}

This analysis of elastic NN scattering data updates our previous
analysis\cite{1} to 1.6~GeV in the laboratory kinetic energy.  The
present analysis extends to 2.5~GeV, which is the limit for elastic
$pp$ differential cross sections measured\cite{2} by the EDDA
collaboration using the cooler synchrotron at COSY.

Measurements with a laboratory kinetic energy near 2 GeV are
particularly interesting as they correspond to a center-of-mass
energy (2.7~GeV) which has been suggested\cite{3} for a broad
dibaryon resonance.  Near this energy, a sharp structure has been
recently reported in the polarization observable $A_{yy}$\cite{4},
and this was taken as support for such a resonance.  A resonancelike
structure, at about the same energy, has also been reported in a
partial-wave analysis by Hoshizaki\cite{5}.  The authors of
Ref.\cite{2} have considered this possibility, but find no evidence
for a resonant excursion in their unpolarized cross sections.
Polarization measurements expected from Saturne II and COSY will
certainly help to clarify this issue.

\section*{THE DATABASE}

Our previous $NN$ scattering analyses\cite{1} were based on about
13,000 $pp$ and about 11,000 $np$ data.  In Ref.\cite{2} the $pp$
analysis extended up to a laboratory kinetic energy of 1.6~GeV; the
$np$ analysis was truncated at 1.3~GeV.  The present database\cite{6}
is considerably larger due both to an expanded energy range for the
$pp$ system and the addition of new data at lower energies.

The database above 1.6 GeV is mainly comprised of cross section
measurements, much of these coming from the EDDA collaboration\cite{2}.
From this source, we have added differential cross sections ranging
from 540~MeV to 2520~MeV in the proton kinetic energy and from
35$^{\circ}$ to 90$^{\circ}$ in the cm scattering angle.  In
constructing the database extension from 1600 to 2500~MeV, we
reexamined a number of references in order to include higher energy
data which had previously been neglected.  This search netted
additional data mainly from ANL (450 points) and Saclay (about 900
points).  The $np$ database has not been increased significantly and,
as a result, we did not extend our analysis of the I~=~0 system.  New
$np$ polarized data have been produced mainly by TRIUMF (100 points).
More details regarding a total database, it can be found in
Refs.\cite{6,7}.

\section*{PARTIAL-WAVE ANALYSIS}

Our first attempts to extend the range of the $NN$ analysis used the
parameterization scheme of Ref.\cite{1}.  These were unsuccessful.
The problem was traced to the basis functions used to expand our
K-matrix elements.  Many of these become nearly degenerate as the
kinetic energy of the incoming nucleon increases to 2.5 GeV.  As a
result, a modified form was used in the present analysis.  Apart from
this difference, the formalism used here is identical to that used in
Ref.\cite{1}.  The reader is directed to Refs.\cite{7,8} for details.

%

Our single-energy and energy-dependent (SM97) results for the dominant
isovector and isoscalar partial-wave amplitudes are displayed in
Fig.~1 and Fig.~2.  (Single-energy analyses were done in order to search 
for structures which may be missing from the energy-dependent fit.)  Here
we also compare with our previous fit (SM94).  In some cases, the
changes are quite large.  This is particularly true near the upper
energy limit of SM94, and for the smaller partial waves.  The effect
of these changes can be clearly seen in Fig.~3, where we show how well
the new EDDA data\cite{2} are reproduced by both SM94 and SM97.  The
influence of this experiment is most pronounced in the forward
direction above 800~MeV.


In general, we find little structure over the higher energy region.
This reflects the smooth, and rather flat, total and reaction cross
sections between 1.5~GeV and 2.5~GeV.  Our fit to these quantities is
displayed in Fig.~4.  Note that the reaction cross sections were
excluded from our fit.  This verifies that the set of total, total
elastic (deduced from differential cross sections), and reaction cross
sections are self-consistent.


The present analysis actually gives an improved fit to the data below
1.6~GeV.  This is due to the altered basis set, found necessary to fit
the higher energy data.  Numerical comparisons are given in Table~1.
Here we see that the COSY data\cite{2} comprise a large fraction of
the total set above 1.6~GeV.  The results of analyses with (SM97) and
without (NM97) this data set show how influential these measurements
have been in determining the amplitudes. (The fits SM97 and NM97 used
identical parameterization schemes. Only the database was changed.)
The COSY data contribute a $\chi^2$/datum of 1.07 from SM97 and 5.6 from 
NM97. The NET $\chi^2$ increased by 3498 when 2121 COSY data were added to 
NM97 and then re-analysed to produce SM97. 

\begin{table}[tbh]
\caption{\it Comparison of present and previous solutions.  Dataset A
             was used in the SM94 analysis\protect\cite{1}.  Dataset B
             contains all data (apart from the EDDA data\protect\cite{2})
             used in generating solution SM97.  See the text for details
             regarding the SM97 and NM97 fits.}
\label{tbl1}
\begin{center}
{\small{
\begin{tabular}{|c|c|c|c|}
 \hline
PWA          & Data        & $\chi^2$/$pp$ data & $\chi^2$/$np$ data  \\
 \hline
             &             &    (0-1600 MeV)    &  (0-1300 MeV)       \\
 \hline
SM94\protect\cite{1}& (dataset A) &     22375/12838    & 17516/10918         \\
SM94\protect\cite{1}& (dataset B) &     22390/12889    & 18480/10843         \\
SM97         & (dataset B) &     20910/12889    & 17400/10843         \\
 \hline
             &             &    (0-2520 MeV)    &  (0-2000 MeV)       \\
 \hline
SM97         & (dataset B) &     26460/14873    & 17440/10854         \\
SM97         & (EDDA dataset\protect\cite{2})   &  2278/2121    & $-$ \\
NM97         & (dataset B) &     25240/14873    & 17280/10854         \\
NM97         & (EDDA dataset\protect\cite{2})   & 11964/2121    & $-$ \\
 \hline
\end{tabular}
}}
\end{center}
\end{table}


\section*{SEARCH FOR LESSER STRUCTURES}

Although our fit (SM97) is parameterized to be devoid of the 
resonance-like structures conjectured to lie around 2~GeV, it is possible 
to add such structures and then look for those observables which are most 
sensitive to such an inclusion.  In Fig.~5 we illustrate how a $^1 S _0$ 
structure at $W _r$ = 2.7~GeV (elastic width is $\Gamma _e$ = 16~MeV, 
inelastic width is $\Gamma _i$ = 67~MeV) would affect $A _{yy}$ and 
$A _{zz}$.

\section*{CONCLUSIONS AND FUTURE PROSPECTS}

We have extended our $pp$ partial-wave analyses nearly 1~GeV beyond
the limit quoted in our previously published results\cite{1}.  The
present range has been selected to include all of the recent elastic
$pp$ cross section data measured by the EDDA group\cite{2}.  We found
that it was possible to simultaneously fit the $pp$ total cross
section data, in particular the precise data of Ref.\cite{9}, along
with differential cross sections from the EDDA collaboration\cite{2}.
The resulting reaction cross sections, which were not included in the
fit, are quite well reproduced.  The predicted reaction cross sections
are consistent with the results of Ref.\cite{10} at lower energies,
but deviate from these and follow the results of Ref.\cite{11} above
1~GeV.
While we find that the partial-wave amplitudes above 1.6~GeV
are smooth and structureless, reflecting the behavior seen in the
total and elastic cross section data, we have also considered the
effect of more localized structures reported in polarization
measurements\cite{3,4}.
As the high energy region was constrained
mainly by cross section data, the present solution should be
considered as a guide to the expected amplitudes.

\section*{acknowledgments}

This work was supported in part by a U.~S. Department of Energy Grant
DE--FG02--97ER41038.  F.~D. was supported by BMBF contract 06HH561Tp2.

\bibliographystyle{unsrt}

\begin{thebibliography}{99}
\bibitem{1} R. A. Arndt {\em {et al.}} Phys.\ Rev.\ {\bf C50}, 2731
(1994).

\bibitem{2} D. Albers {\em {et al.}} Phys.\ Rev.\ Lett. {\bf 78},
1652 (1997).

\bibitem{3} A recent study was made by E. L. Lomon, Few-Body Systems,
Suppl. {\bf 7}, 213 (1994).  For a different approach to this problem
see B. Z. Kopeliovich and F. Niedermayer, Phys.\ Lett.\ {\bf B117},
101 (1982), and S. J. Brodsky and G. F. de Teramond, Phys.\ Rev.\
Lett.\ {\bf 60}, 1924 (1988).  A review is given by I. I. Strakovsky,
J.\ of Part.\ and Nuclei,\ {\bf 22}, 296 (1991).

\bibitem{4} J. Ball {\em {et al.}} Phys.\ Lett.\ {\bf B320}, 206
(1994).

\bibitem{5} N. Hoshizaki in: {\it {Proc. of the Workshop on the
Experiments by Polarized Proton and Electron Beams, Tsukuba, Japan,
1988}}, p.~201.

\bibitem{6} These results and corresponding databases can be viewed
and compared with previous analyses at the {\it WWW} site {\bf
http://clsaid.phys.vt.edu} or through a {\it Telnet} call to
{\bf clsaid.phys.vt.edu} with {\it user}: {\bf said} (no password).

\bibitem{7} R. A. Arndt {\em {et al.}} Submitted to Phys.\ Rev.\
{\bf C}, E-Print Archive: NUCL--TH/9706003.

\bibitem{8} R. A. Arndt {\em {et al.}} Phys.\ Rev.\ {\bf D35}, 128
(1987), M. H. MacGregor {\em {et al.}} Phys.\ Rev.\ {\bf 169}, 1128
(1968).

\bibitem{9} D. V. Bugg {\em {et al.}} Phys.\ Rev.\ {\bf 146}, 980
(1966).

\bibitem{10} B. J. VerWest and R. A. Arndt, Phys.\ Rev.\ {\bf C25},
1979 (1982).

\bibitem{11} J. Bystricky {\em {et al.}} J.\ Physique {\bf 48}, 1901
(1987).

%
%

\end{thebibliography}

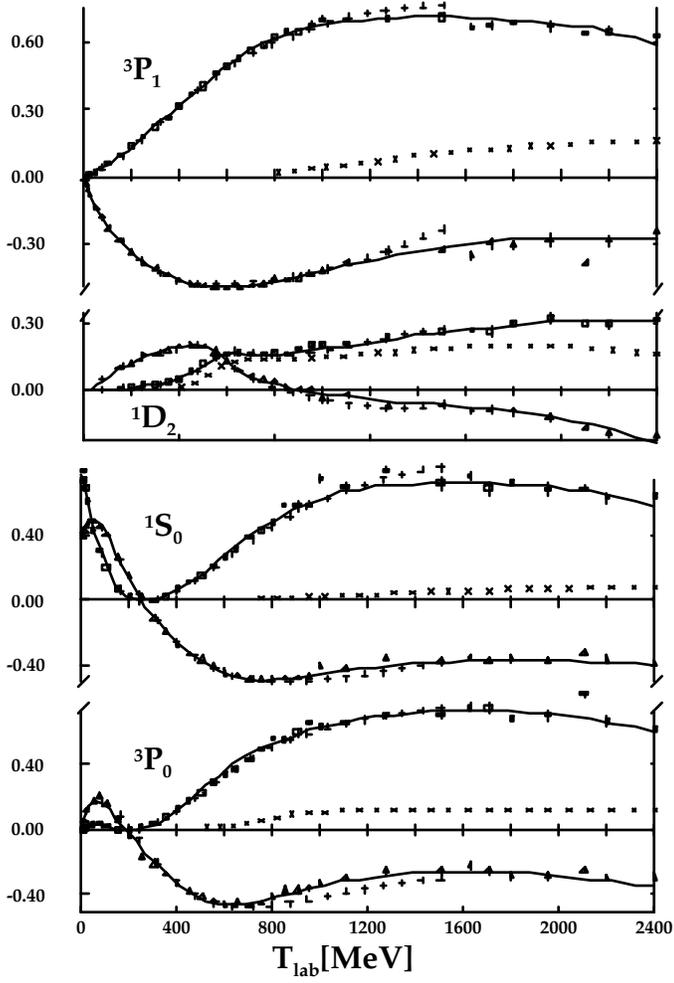
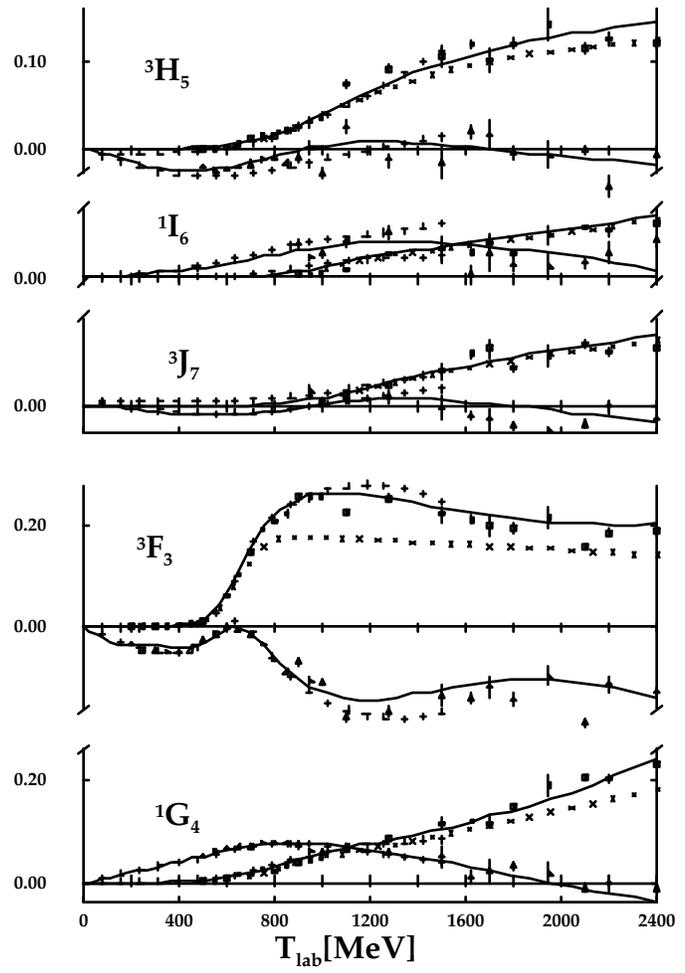
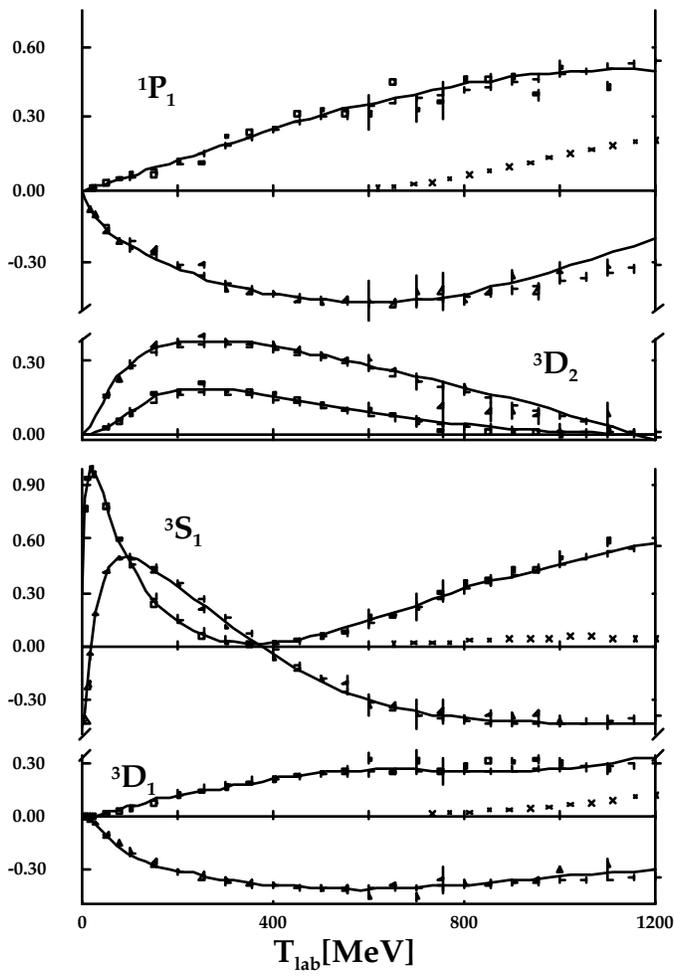
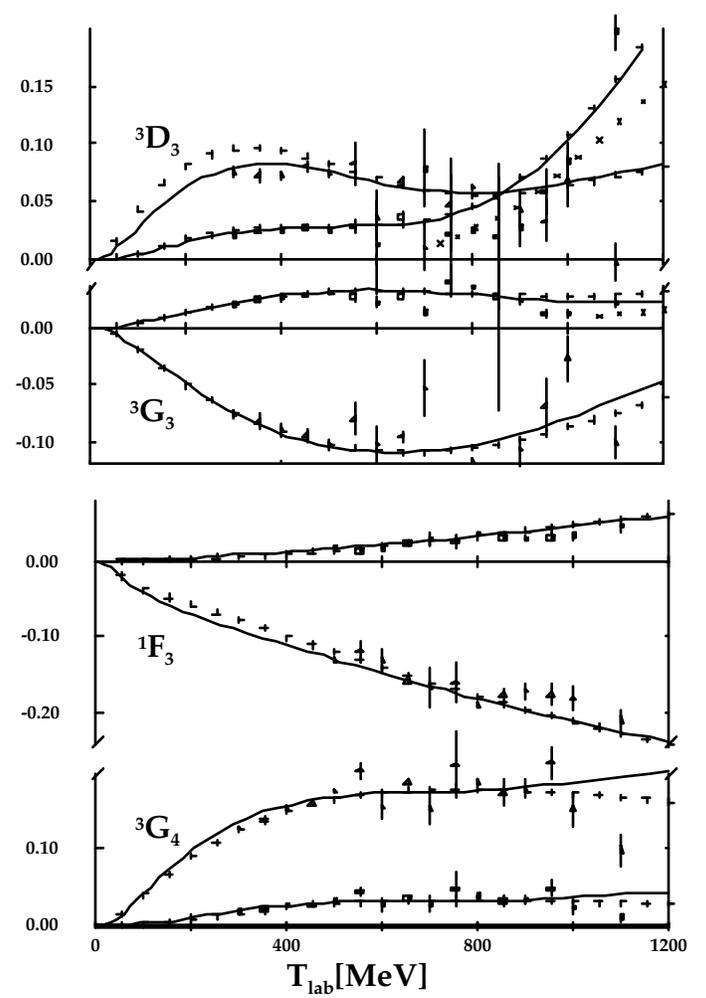
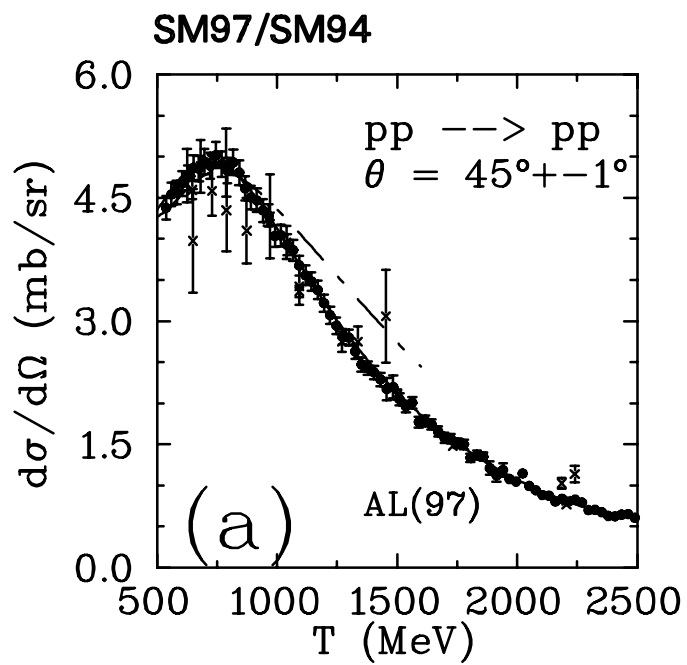
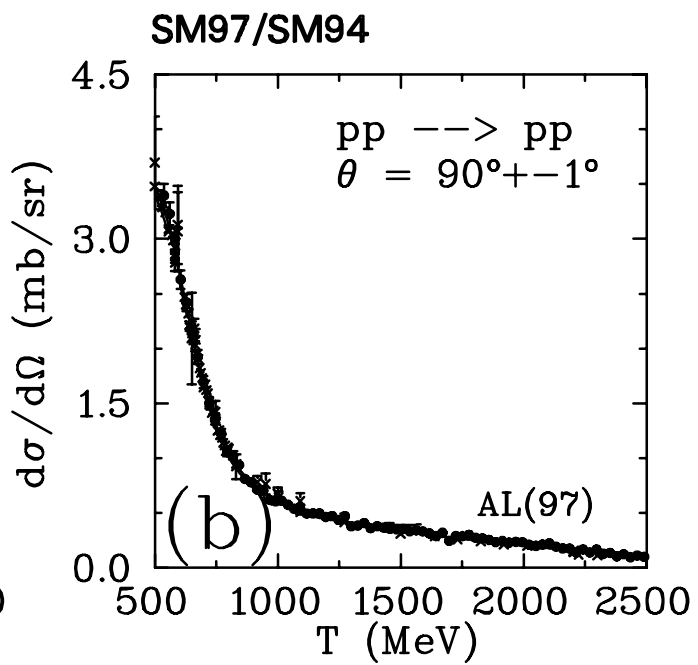
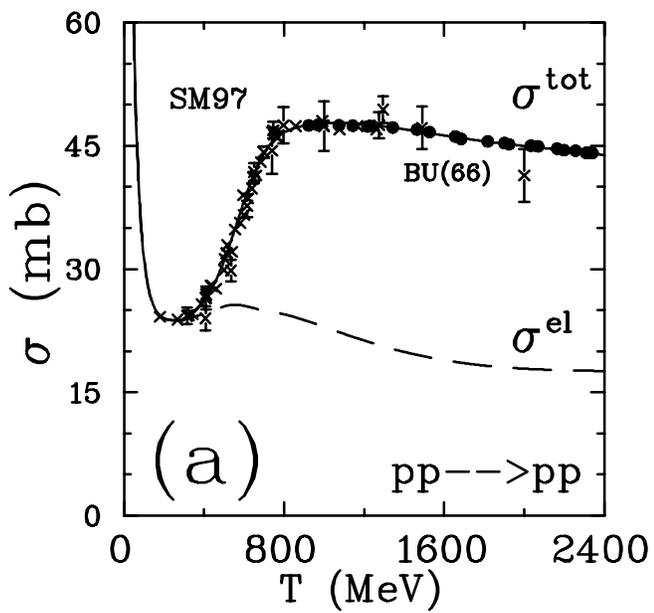
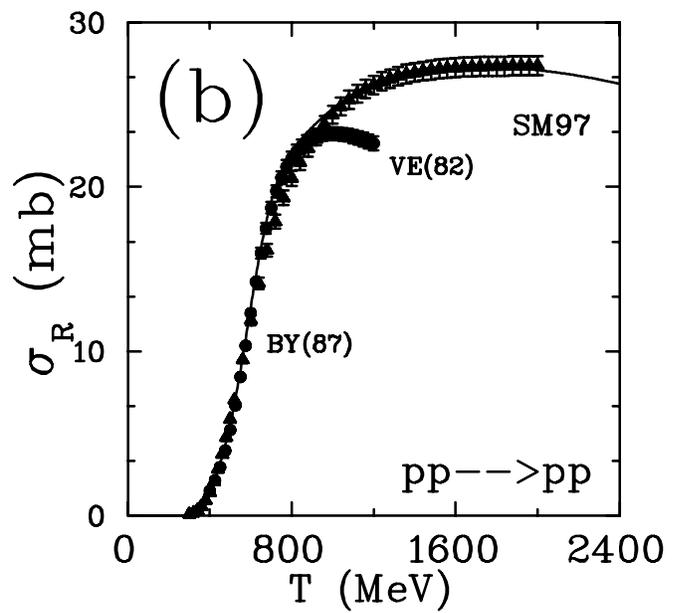
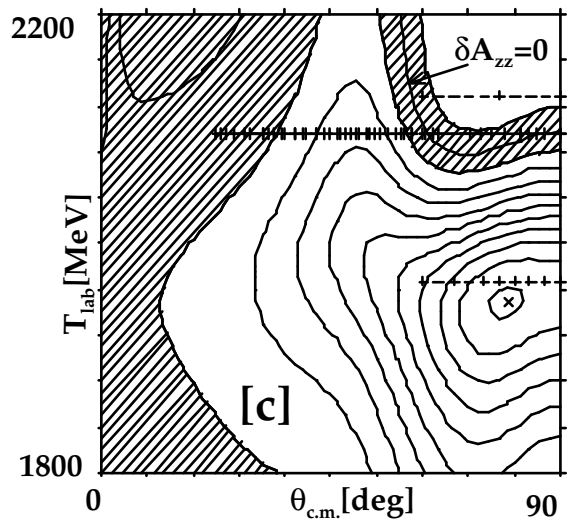
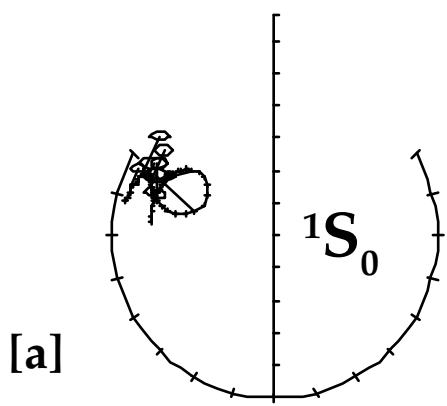
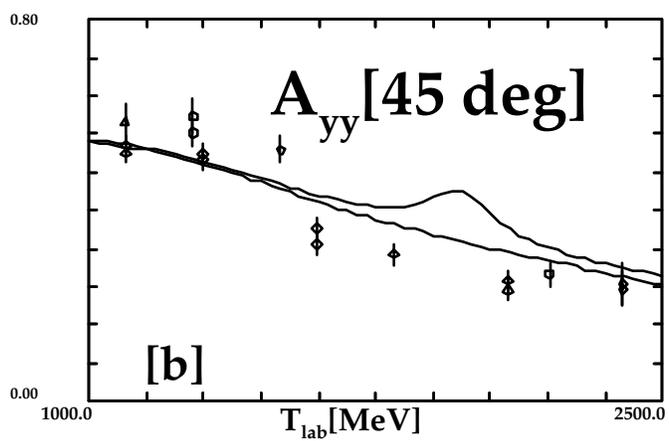
\end{document}